# CHARACTERISTICS OF THE SECONDARY ELECTRONS CALIBRATION BEAM OF THE ACCELERATOR S-25R «PAKHRA»

V.I. Alekseev[a], V.A. Baskov[a]*, V.A. Dronov[a], A.I. L'vov[a], A.V. Koltsov[a], Yu.F. Krechetov[a], V.V. Polyansky[a], S.S. Sidorin[a]

[a] *P.N. Lebedev Physical Institute, Moscow, 119991 Russia*

[b] *Joint Institute for Nuclear Research, Dubna, Moscow Region, 141980 Russia*

*E-mail: baskov@x4u.lebedev.ru

The characteristics of the secondary electrons` calibration quasi-monochromatic beam of the accelerator S-25R «Pakhra» of the Lebedev Physical Institute of the Russian Academy of Sciences (LPI) on the basis of magnet SP-57 are presented. With an electron energy in the range of 45-280 MeV, a collimator diameter in front of the trigger counters of 3 mm and copper Converter thicknesses of 1-3 mm, the energy resolution and beam intensity were $\sigma$ =4.4-2.2 % and ~16 e /sec, respectively.

In nuclear physics research, an important stage in the creation of experimental facilities is the research of the equipment and detectors` characteristics (energy and time resolution, efficiency, etc.) [1-3]. A quasi-monochromatic beam of secondary electrons was created on the basis of the SP-57 magnet to perform calibration work on the electronic accelerator S-25R «Pakhra» of the Lebedev Physical Institute of the Russian Academy of Sciences (LPI) with the energy of accelerated electrons in the range 200÷850 MeV and beam intensity ~2·$10^{12}$ e−/sec on the basis [4]. The relative energy resolution of the calibration beam in the electron energy range E = 98-294 MeV was $\sigma$ = 10-4.5 %, respectively $\sigma$ = $\Delta E/E/2.35$, $\Delta E$-the full width of the energy spectrum of the electron beam at half its height, E-the average energy of the electron beam).

In the process of calibration beam operation it becames clear that for calibrations of a number of detectors the received accuracy is not enough. Research was carried out in order to find out the influence of the main elements forming the beam on the energy resolution – the thickness of the copper converter $t_c$ and the diameter of the input collimator *d* in front of the trigger counters. This paper presents the results of research.

The scheme of the calibration beam presented in Fig. 1. The electrons resulting from interaction of the photon beam with the converter (3) were deflected into a lead collimator K (6) with an inlet diameter of 10 mm and a thickness of 7 cm, located at the angle of $\varphi = 36°$ relative to the primary photon trajectory at a distance of 3 m from the poles of the magnet.

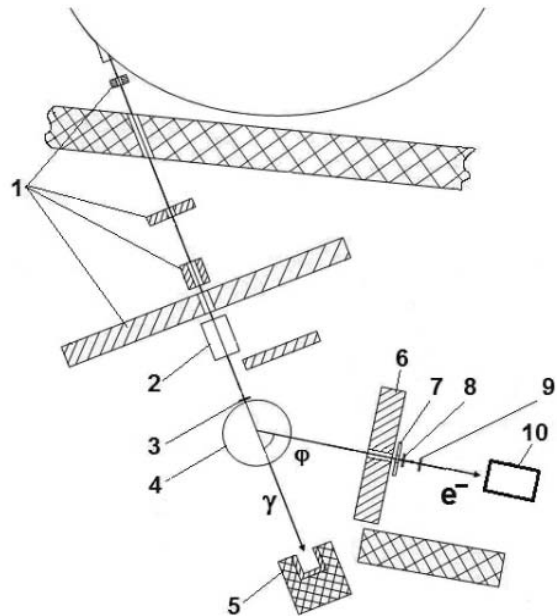

**Fig. 1.** Scheme of quasi-monochromatic beam of secondary electrons of «Pakhra» accelerator: 1 - lead collimators; 2-cleaning magnet SP-3; 3 - converter; 4 - magnet SP-57; 5- photon beam absorber («mortuary»); 6 - collimator; 7 - scintillation counter anticoincidence A; 8, 9-trigger scintillation counters $S_1$ and $S_2$; 10- total absorption Cherenkov spectrometer (TAChS).

The trigger signal T was formed by from the coincidence of the polystyrene scintillation counters $S_1$, $S_2$ and the counter of anticoincidence A with an orifice of ∅10 mm (T=( $S_1 \cdot S_2$)·A). The counters $S_1$ and $S_2$ were size of $10 \times 10 \times 5$ mm$^3$, counter A-$60 \times 90 \times 10$ mm$^3$. The distance between $S_1$ and $S_2$ was 32 cm. The photoelectron multipliers PMT-85 with the voltage across the power divider U = 1000 V were used in the counters.

The energy of secondary electrons was determined by a total absorption Cherenkov spectrometer (TAChS) [4].

The characteristics of a quasi-monochromatic electron beam were determined by the spectrum difference method with and without converter. At the first stage, the main characteristics of the electron beam, which were the energy magnitude and energy resolution, were studied. At the second stage, when only air was the converter, the same characteristics were studied. Then, the spectra obtained with a complex converter (copper+air) were subtracted from the spectra obtained with an air converter [4].

The research of the energy resolution of the electron beam was carried out stepwise. At the first stage, the energy resolution of the beam with an orifice of collimator (6) of 10 mm was studied, followed by a change in the thickness of the converter from 1 mm to 3 mm. At stages 2 and 3, the diameter of the input collimator was consistently changed by 5 mm and 3 mm, also with a further change in the thickness of the converter from 1 mm to 3 mm.

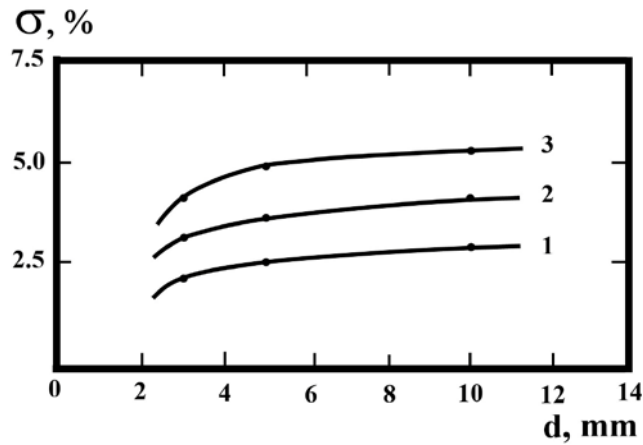

**Fig. 2.** The dependence of the energy resolution of the secondary electron beam on the diameter of the collimator orifice d and the thickness of the converter $t_c$ at the electron beam energy E = 280 MeV. In the picture: 1- $t_c$ = 3 mm; 2- $t_c$ = 5 mm; 3- $t_c$ = 10 mm.

Figure 2 shows the energy resolution of the secondary electron beam depending on the thickness of the converter (3) and the diameter of the collimator (6) at the electron energy E = 280 MeV. It is seen that the energy resolution of the electron beam improves with decreasing thickness of the converter and the diameter of the collimator orifice. The lowest relative energy resolution of the beam is σ = 2.2 % with the thickness of the copper converter $t_c$ = 1 mm and the diameter of the collimator orifice d = 3 mm.

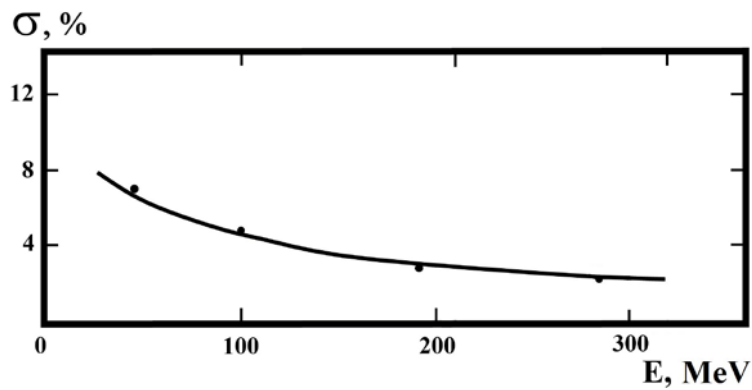

**Fig. 3.** The dependence of the energy resolution of the secondary electron beam on the energy at $t_c$ = 1 mm and d = 3 mm.

The dependence of the energy resolution of the secondary electron beam on the energy at $t_c = 1$ mm and $d = 3$ mm is presented in Fig. 3. The worst energy resolution of the electron beam reaches $\sigma = 7$ % at the lowest studied energy $E = 45$ MeV.

The dependence of the electron beam intensity on the thickness of the converter (3) and the diameter of the collimator orifice (6) at the electron beam energy $E = 280$ MeV is presented in Fig. 4. The intensity increases with increase in the thickness of the converter and the diameter of the collimator orifice, while the relative energy resolution deteriorates (Fig. 2). The maximum beam intensity is I≈ 53 e⁻/sec at $t_c = 3$ mm and $d = 10$ mm.

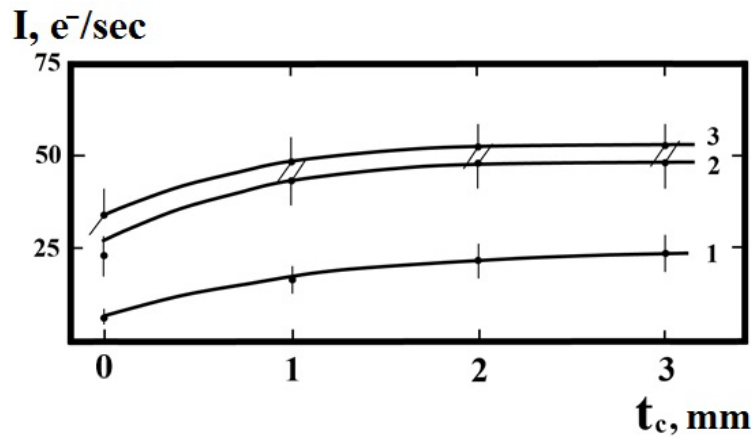

**Fig. 4.** The dependence of the electron beam intensity on the thickness of the converter $t_c$ and the diameter of the collimator orifice d at the electron beam energy E = 280 MeV. In the figure: 1-d = 3 mm; 2-d =5 mm; 3-d = 10 mm.

The dependence of the electron beam intensity on the energy at $t_c = 1$ mm and $d = 3$ mm is illustrated in Fig. 5. The intensity reaches a maximum of about 16 e⁻/ sec at energies greater than 80 MeV.

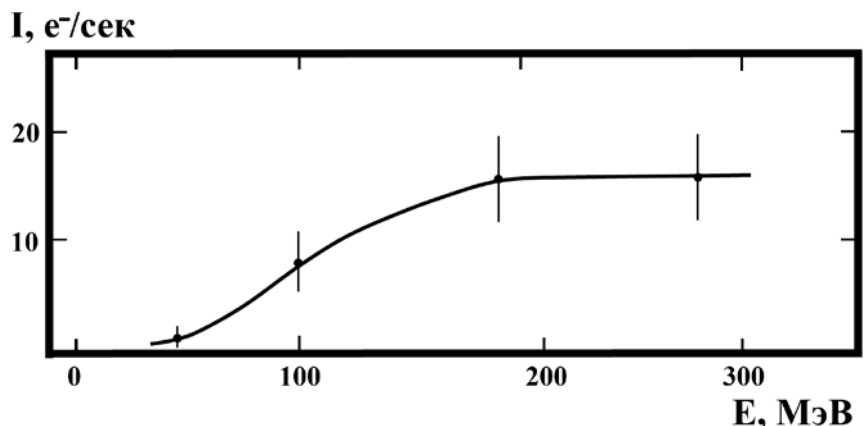

**Fig. 5.** The dependence of the electron beam intensity on energy at $t_c = 1$ mm and $d = 3$ mm.

In figures 4 and 5, air is the converter at $t_c = 0$ in the open interval between the cleaning magnet SP-3 (2) and the spectrometric magnet SP-57 (4).

Thus, the best energy resolution of the calibration beam of secondary electrons is $\sigma = 2.2$ % with the thickness of the copper converter $t = 1$ mm, the diameter of the collimator orifice $d = 3$ mm and the beam energy $E = 280$ MeV. At the same time the intensity of the electron beam is ~16 $e^-$/ sec.

This work was supported by grants from the Russian Foundation for Basic Research (NICA - RFBR) No. 18-02-40061 and No. 18-02-40079.